\makeatletter
\newcommand{\dontusepackage}[2][]{%
  \@namedef{ver@#2.sty}{9999/12/31}%
  \@namedef{opt@#2.sty}{#1}}
\makeatother
\dontusepackage{subfigure}

\documentclass[]{article}

\usepackage{lmodern}
\usepackage{amssymb,amsmath}
\usepackage{ifxetex,ifluatex}
\usepackage[usenames,dvipsnames]{color}
\usepackage{fixltx2e} 
\ifnum 0\ifxetex 1\fi\ifluatex 1\fi=0 
  \usepackage[T1]{fontenc}
  \usepackage[utf8]{inputenc}
\else 
  \ifxetex
    \usepackage{mathspec}
    \usepackage{xltxtra,xunicode}
  \else
    \usepackage{fontspec}
  \fi
  \defaultfontfeatures{Mapping=tex-text,Scale=MatchLowercase}
  
\fi
\IfFileExists{upquote.sty}{\usepackage{upquote}}{}
\IfFileExists{microtype.sty}{%
\usepackage{microtype}
\UseMicrotypeSet[protrusion]{basicmath} 
}{}
\usepackage[margin=1.0in,bottom=1.5in]{geometry}
\usepackage[square,numbers,sort&compress]{natbib}
\usepackage{listings}
\usepackage{longtable,booktabs}
\usepackage{graphicx}
\makeatletter
\def\maxwidth{\ifdim\Gin@nat@width>\linewidth\linewidth\else\Gin@nat@width\fi}
\def\maxheight{\ifdim\Gin@nat@height>\textheight\textheight\else\Gin@nat@height\fi}
\makeatother
\setkeys{Gin}{width=\maxwidth,height=\maxheight,keepaspectratio}
\usepackage{caption}
\usepackage{float}
\ifxetex
  \usepackage[setpagesize=false, 
              unicode=false, 
              xetex]{hyperref}
\else
  \usepackage[unicode=true]{hyperref}
\fi
\hypersetup{breaklinks=true,
            bookmarks=true,
            pdfauthor={},
            pdftitle={Enabling wave-based inversion on GPUs with randomized trace estimation},
            colorlinks=true,
            citecolor=black,
            urlcolor=blue,
            linkcolor=black,
            pdfborder={0 0 0}}
\usepackage[preprint]{neurips_2019}
\usepackage{url}
\usepackage{booktabs}
\usepackage{amsfonts}
\usepackage{nicefrac}
\usepackage{microtype}

\title{Enabling wave-based inversion on GPUs with randomized trace estimation}
\author{Mathias Louboutin\textsuperscript{1}, Felix J.
Herrmann\textsuperscript{1,2}\\\textsuperscript{1} School of Earth and
Atmospheric Sciences, Georgia Institute of
Technology\\\textsuperscript{2} School of Computational Science and
Engineering, Georgia Institute of Technology\\}
\date{}

\begin{document}
\maketitle

\section{Summary}\label{summary}

By building on recent advances in the use of randomized trace estimation
to drastically reduce the memory footprint of adjoint-state methods, we
present and validate an imaging approach that can be executed
exclusively on accelerators. Results obtained on field-realistic
synthetic datasets, which include salt and anisotropy, show that our
method produces high-fidelity images. These findings open the enticing
perspective of 3D wave-based inversion technology with a memory
footprint that matches the hardware and that runs exclusively on
clusters of GPUs without the undesirable need to offload certain tasks
to CPUs.

\section{Introduction}\label{introduction}

Subsurface imaging has recently thrived building on advances in
wave-equation based methods such as Full-Waveform Inversion (FWI) and
Reverse-Time Migration (RTM) \citep{tarantola, virieux}. However, these
methods rely on extremely high computational and memory costs, which
explains the relative limited widespread adaptation of these
technologies. Unfortunately, exceedingly large memory footprints are
inherent to the adjoint-state method \citep{lionsjl1971, tarantola},
which requires storage (in memory, on disk, possibly compressed) of the
complete time history of the forward modeled wavefield in order to
compute the imaging condition that correlates this forward wavefield
with the time-reversed solution of the adjoint wave equation. Because
saving forward modeled wavefields requires terabytes of memory for
industry scale high-frequency 3D imaging, memory usage has been and
continues to be a major bottleneck on standard with the exception
perhaps of dedicated high-memory nodes available in the cloud. While
dedicated nodes relief some of the memory pressure, they do not allow
use of accelerators to speed up computations. Contrary to conventional
computer hardware, memory on accelerators comes at a premium, which is
problematic given the large memory footprint of adjoint-state methods.
To address this problem, several methods have been proposed over the
years where excessive memory footprints are offset by incurring
computational overhead. A good example of such an approach is the method
of optimal checkpointing proposed by \citet{Griewank} and
\citet{Symes2007}. This method was initially introduced to tackle memory
limitation of CPUs and has been used successfully in 3D seismic
applications. To further limit the computational overhead,
\citet{kukrejacomp} recently supplemented this approach by adding
on-the-fly compression and decompression of the forward wavefields. In
situations where the wave physics is reversible, researchers
\citep{McMechan, Mittet, RaknesR45} have shown that forward wavefields
can also be recomputed from boundary values. Unfortunately, both
approaches are challenged by underlying assumptions. They also require a
relative high of algorithmic complexity. This explains why GPUS-native
implementations of adjoint-state methods including on-the-fly
compression remain illusive.

While recomputing forward wavefields as part of memory-footprint
mitigation certainly has its merits, there exist simpler randomized
approaches where memory use is traded against computational overhead and
controllable error. Unlike approaches that aim to compute gradients
exactly, these methods approximate the gradient with the aim to reduced
computational and memory costs at the expense of a controllable loss in
accuracy. Examples of such an approach include to working with random
subsets of shots \citep{friedlander} or simultaneous shots
\citep{romero, krebs, Moghaddamsefwi, haberse, leeuwenfwi}, or with
randomized singular value decompositions
\citep{vanleeuwen2015GPWEMVA, yang2020lrpo} and random-trace estimation.
The latter was used by \citet{haberse} to analyze computational speedups
of full-waveform with computational simultaneous sources. As long as the
errors are controlled, these methods lead to equivalent inversion
results at fraction of the computational costs (e.g.~speedups of a
factors of seven have been reported for FWI \citep{krebsse}). Following
ideas from randomized linear algebra to estimate the trace of a matrix,
\citet{louboutin2021SEGulm} proposed an approximation of the
adjoint-state method that leads to major memory improvements and is
relatively easy to implement and supported by theory
\citep{Avron, hutchpp}, guaranting convergence including bounds on the
accuracy. However, unlike other approximate methods, such as on-the-fly
Fourier-based \citep{witte2018cls} or lossy compression-based
algorithms\citep{kukrejacomp}, the artifacts introduced by the proposed
randomized trace estimation are incoherent and appear as Gaussian-like
noise, which can be handled easily by sparsity-promoting imaging
\citep{witte2018cls}. While the initial results of the randomized trace
estimation on a simple 2D synthetic were encouraging
\citep{louboutin2021SEGulm}, we submit the proposed approximation to
additional scrutiny by considering complex imaging examples that involve
salt (SEAM model \citep{segseamI}) and anisotropy \citep{Thomsen1986}
(BP TTI model).

Our contributions are organized as follows. First, we briefly introduce
randomized trace estimation and its computational benefits during RTM
and the formation of horizontal subsurface-offset image volumes
\citep{symesiv}. Next, we show its application to two representative
examples, long-offset subsalt imaging on the 2D SEAM acoustic model with
a sparse ocean-bottom node acquisition, and TTI anisotropic imaging of
the 2D BP TTI model. With these examples, we validate the computational
efficiency and practicality of our method on GPUs available on Azure
(the \texttt{Standard\_NC6} virtual machine).

\section{Methodology}\label{methodology}

Before we demonstrate the advocacy of the proposed methodology on
complex imaging problems, let us first quickly discuss how randomized
trace estimation can be used to reduce the memory footprint of
adjoint-state wave-based seismic imaging. We do this by showing that
applying the imaging condition corresponds to computing the trace of a
matrix.

\subsection{Randomized trace
estimation}\label{randomized-trace-estimation}

Approximating \citep[Avron;][]{hutchpp} the identity $\mathbf{I}$ by
\begin{equation}
\operatorname{tr}(\mathbf{A}) =\operatorname{tr}\left(\mathbf{A} \mathbb{E}\left[\mathbf{z} \mathbf{z}^{\top}\right]\right) =\mathbb{E}\left[\operatorname{tr}\left(\mathbf{A} \mathbf{z} \mathbf{z}^{\top}\right)\right]
= \mathbb{E}\left[\mathbf{z}^{\top} \mathbf{A} \mathbf{z}\right] \approx \frac{1}{r}\sum_{i=1}^r\left[\mathbf{z}^{\top}_i \mathbf{A} \mathbf{z}_i\right]= \frac{1}{r}\operatorname{tr}\left(\mathbf{Z}^\top \mathbf{A}\mathbf{Z}\right)
\label{randomtrace}
\end{equation}
 lies at the heart of randomized trace estimation where the
$\mathbf{z}_i$'s are random probing vectors for which
$\mathbb{E}(\mathbf{z}^{\top}\mathbf{z})=1$ with $\mathbb{E}$ is the
stochastic expectation operator. This above estimator for the trace (sum
of the diagonal of the matrix $\mathbf{A}$,
$\operatorname{tr}(A)=\sum_i \mathbf{A}_{ii}$) is unbiased (i.e., exact
in expectation) and converges to the true trace with an error that
decays with $r$ and without access to the entries of $\mathbf{A}$. Only
actions of $\mathbf{A}$ on the probing vectors are needed and we exploit
this property and the specific structure of the matrix $\mathbf{A}$ in
gradient calculations for wave-equation based inversion. Motivated by
recent work \citep{hutchpp, graff2017SINBADFlrp} we also employ a
partial \texttt{qr} factorization \citep{trefethen1997numerical} that
approximates the range of the matrix $\mathbf{A}$---i.e., we approximate
the trace with probing vectors
$\begin{bmatrix}\mathbf{Q},\thicksim\end{bmatrix} = \operatorname{qr}(\mathbf{A}\mathbf{Z})$
where $\mathbf{Z}$ is a Rademacher random matrix of $\pm 1$.

\subsection{Approximate gradient
calculations}\label{approximate-gradient-calculations}

While the presented randomized approach carries over to arbitrary
complex wave physics, we derive our memory reduced gradient
approximation for the isotropic acoustic case where the gradient for a
single source $\delta\mathbf{m}$ can be written as
\begin{equation}
\delta\mathbf{m} = \sum_{t=1}^{n_t} \mathbf{\ddot{u}}[t] \mathbf{v}[t].
\label{iccc}
\end{equation}
 In this expression, $\mathbf{u}[t], \mathbf{v}[t]$ are the vectorized
(along space) full-space forward and adjoint wavefields at time index
$t=1\cdots n_t$ with $n_t$ the number of timesteps. The symbol $\ddot{}$
represents second-order time derivative. To arrive at a form where
randomized trace estimation can be used, we write the above zero-lag
imaging condition over time as the trace of the outer product for each
space index $\mathbf{x}$ separately. By using the dot product property,
$\sum \mathbf{x}_i \mathbf{y}_i=\mathbf{x}^\top\mathbf{y}=\operatorname{tr}(\mathbf{x}\mathbf{y}^\top)$,
and the approximation in Equation~\ref{randomtrace}, the gradient at
location $\mathrm{x}$ becomes
\begin{equation}
\begin{split}
\delta\mathbf{m}[\mathbf{x}] = \operatorname{tr}\left(\mathbf{\ddot{u}}[\cdot, \mathbf{x}]\mathbf{v}[\cdot, \mathbf{x}]^\top\right) \approx \frac{1}{r} \operatorname{tr}\left(\bar{\ddot{\mathbf{u}}}[\cdot, \mathbf{x}]\bar{\mathbf{v}}[\cdot, \mathbf{x}]^\top\right)\quad\text{with}\quad \bar{\ddot{\mathbf{u}}}=\mathbf{Q}^\top \mathbf{\ddot{u}}, \bar{\mathbf{v}}=\mathbf{Q}^\top \mathbf{v}. \\
\end{split}
\label{optr}
\end{equation}
 Contrary to the sum over all timesteps $t$, the zero-offset imaging
condition in Equation~\ref{optr} involves storage of the compressed
wavefields for only $r\ll n_t$ timesteps. This compression not only
significantly reduces the memory footprint but it also lessens the
computational cost of computing imaging conditions as a function of the
horizontal subsurface offset
\begin{equation}
\delta\mathcal{M}[\mathbf{x},\mathbf{h}] \approx \frac{1}{r} \operatorname{tr}\left(\bar{\ddot{\mathbf{u}}}[\cdot, \mathbf{x}+\mathbf{h}]\bar{\mathbf{v}}[\cdot, \mathbf{x}-\mathbf{h}]^\top\right)
\label{ssoffset}
\end{equation}
 with $\mathbf{h}$ the horizontal subsurface offset and
$\delta\mathcal{M}[\mathbf{h}]$ the subsurface image volume. As with
computing the zero-offset imaging condition in Equation~\ref{optr}, the
cost of computing these volumes is also reduced by a factors of $r/n_t$.
Also notice that
$\delta\mathbf{m}[\mathbf{x}]=\delta\mathcal{M}[\mathbf{x},h]\vert_{\mathbf{h}=0}$.

\begin{figure}
\centering
\includegraphics[width=1.000\hsize]{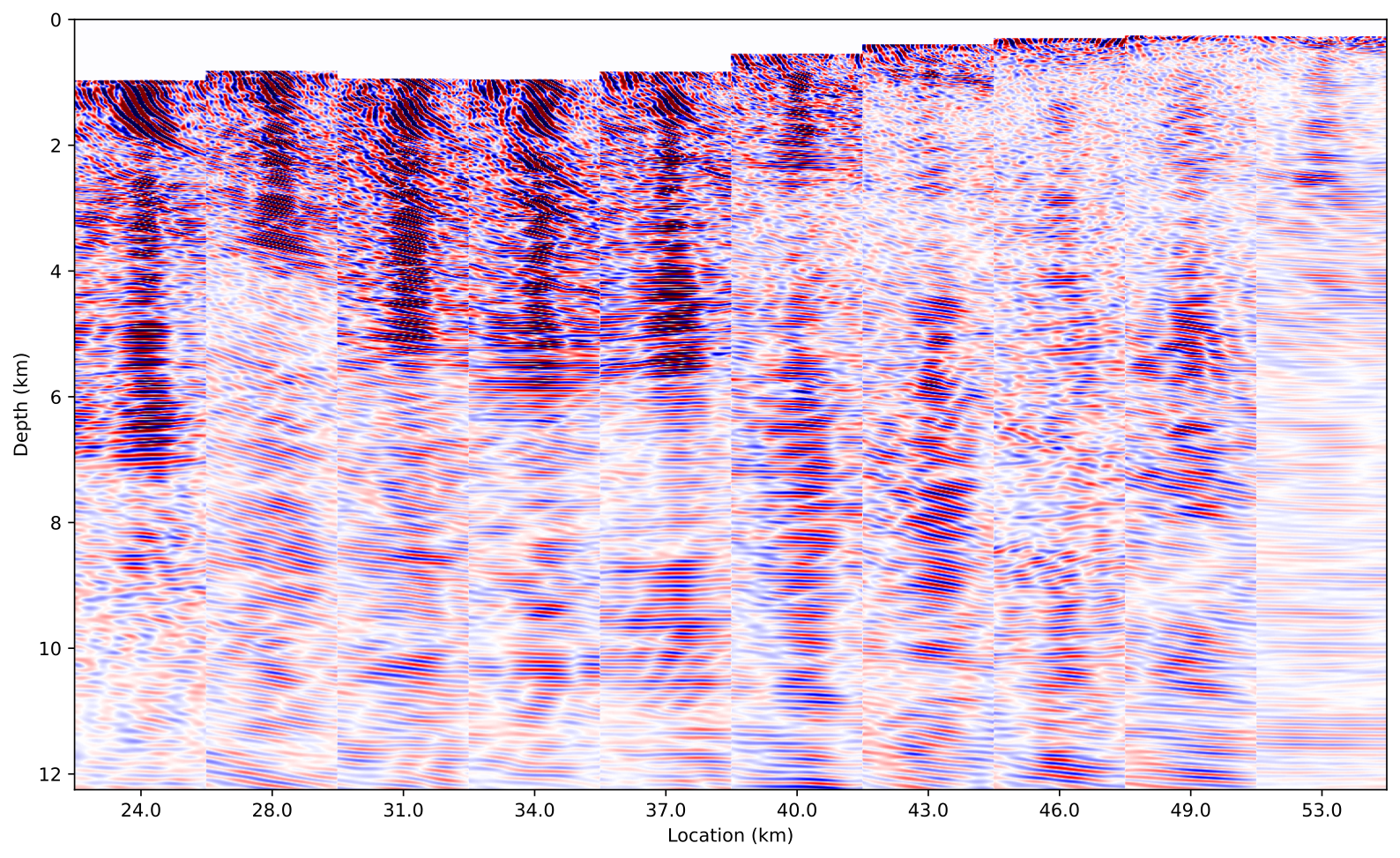}
\caption{Subsurface offset gathers with a \texttt{-500m} to
\texttt{500m} horizontal offset.}\label{fig:2d-bp-sso}
\end{figure}

Figure~\ref{fig:2d-bp-sso} contains subsurface offset gathers for the
2007 BP TTI model discussed in more detail below. Despite the fact that
we used only a limited ($r=64\ll n_t$) probing factors, the images
gathers are properly focused around and nearly noise free thanks to the
noise stacking. Each image volume is using $81$ offsets
($-500m:12.5m:500m$) and is effectively using more memory that we needed
to store the compressed wavefields needed to compute these volumes,
highlighting how memory frugal the proposed trace estimation method
realy is.

\section{Synthetic case studies}\label{synthetic-case-studies}

To validate the proposed technology, we consider the 2D acoustic SEAM
model and the 2D 2007 BP TTI model. We chose these models because they
are complex and in need of a large number of timesteps. As the examples
included below, we obtain reasonable RTM images for a limited number of
probing vectors ($r=64$), yielding a memory reduction by a factor of
$100\times$---$150\times$ reducing the memory footprint to less than
2Gb. This drastic memory reduction allows us to fully take advantage of
accelerators by performing the RTM imaging natively on GPUs without
relying on advanced IO or checkpointng techniques. More details on
memory gains are included in Table~\ref{memrtm}. We refer to
\citep{louboutin2021SEGulm} for more details on memory cost and
computational overhead of the proposed method.

\begin{table}
\centering
\begin{tabular}{cccc}
\toprule\addlinespace
& Standard RTM & Trace Estimation & Gain ($\times$)\tabularnewline
\midrule
SEAM & 380/94 Gb & 1.4/.7 Gb & 271--134\tabularnewline
BP TTI & 337/75 Gb & 2.1/1.05 Gb & 160--71\tabularnewline
\bottomrule
\end{tabular}
\caption{Memory usage of standard adjoint-state RTM (full
history/sampled at \texttt{4ms}) versus imaging via random trace
estimation with $r=64/32$ on the 2D SEAM model and 2D BP TTI
model.}\label{memrtm}
\end{table}

\subsection{2D SEAM model}\label{d-seam-model}

\begin{figure}
\centering
\includegraphics[width=1.000\hsize]{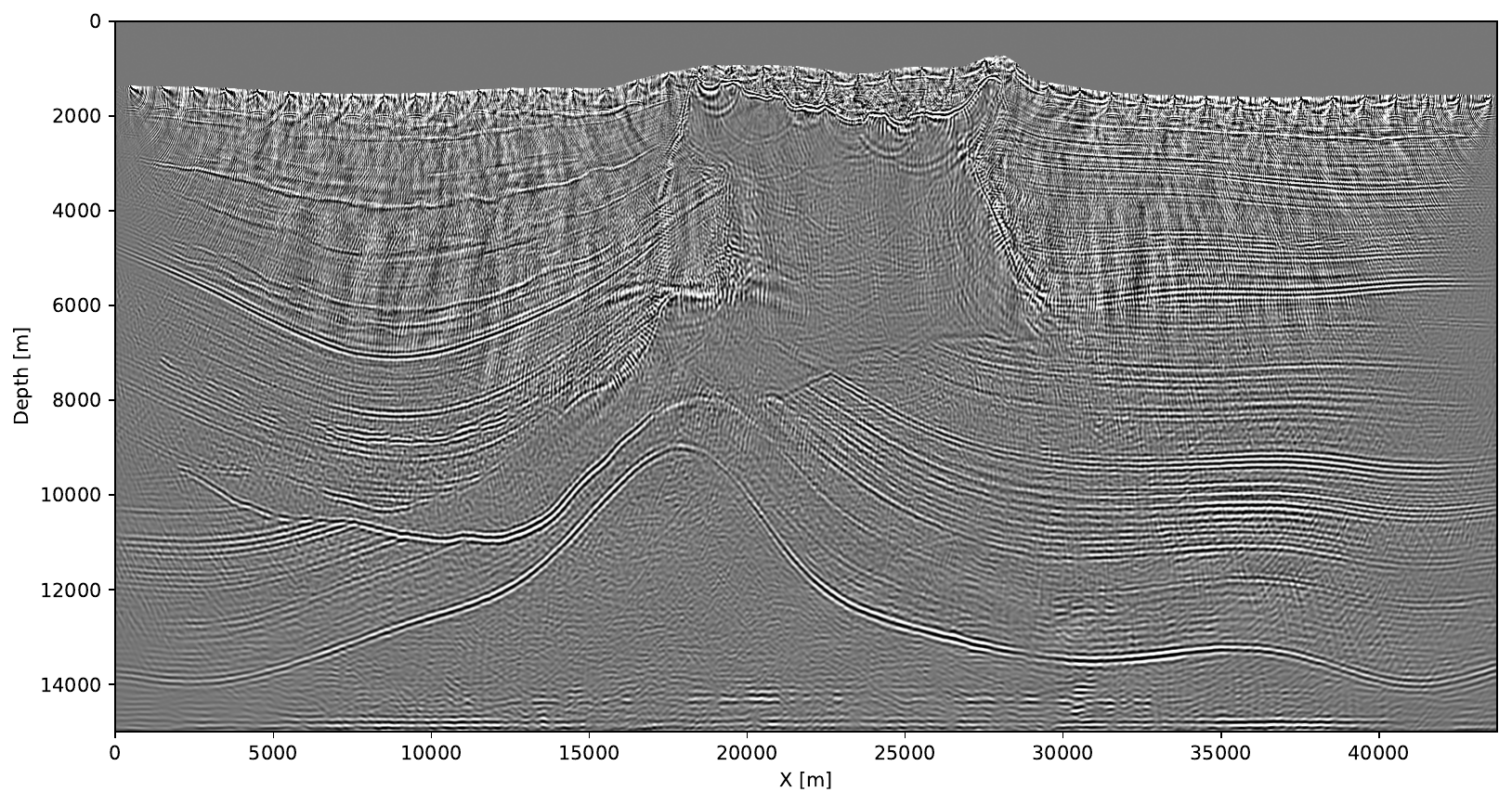}
\caption{2D RTM on the SEAM model with sparse OBN
acquisiton.}\label{fig:2d-seam}
\end{figure}

To study the behavior of our approximation on long-offset sparse OBN
acquisition, we consider a 2D slice of the SEAM salt model
\citep{segseamI}. Because this type of acquisition improves the
illumination of large salt bodies in the Gulf of Mexico, this type of
acquisition has recently gained in popularity. The survey consists of
$44$ OBNs one kilometer apart. At the surface, sources are located every
$12.5m$ at a depth of $12.5m$. We idealized this dataset by modeling
with reciprocity, which leads to $44$ densely sampled common receiver
gathers that serve as input to our approximate imaging approach. As we
can clearly see from Figure~\ref{fig:2d-seam}, we are able to produce a
good image despite complexity of the model and drastic compression of
the wavefield (we use only $r=64$ probing vectors. Even though we incur
limited noise mostly in shallow areas, we argue that these noisy
artifacts can easily be removed. At greater depth, the imprint of the
noise is less leading to good resolution below the salt.

\subsection{2007 BP TTI}\label{bp-tti}

\begin{figure}
\centering
\includegraphics[width=1.000\hsize]{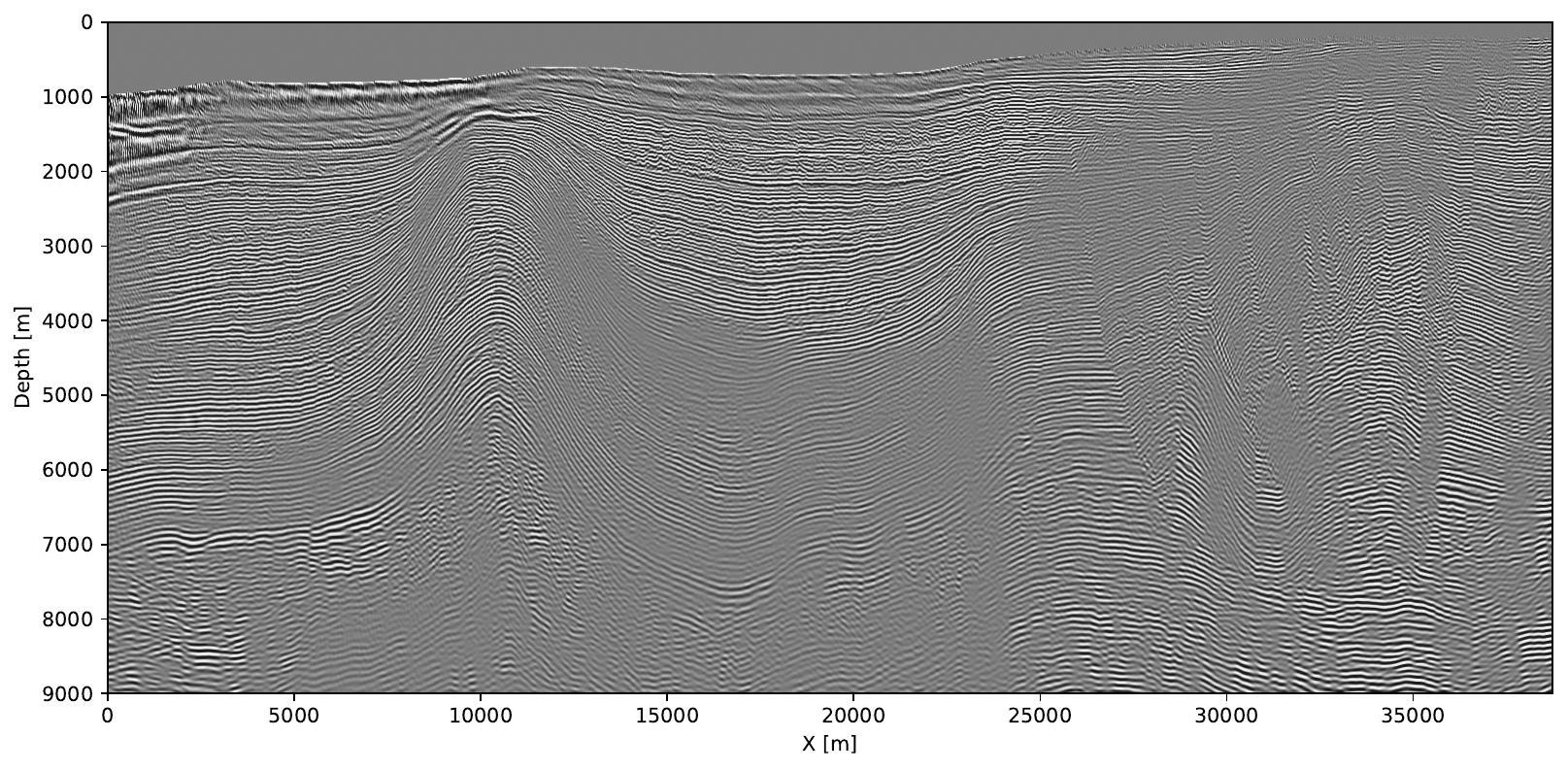}
\caption{2D RTM on the 2007 BP TTI model with a marine acquisition.}
\label{fig:2d-bp}
\end{figure}

To demonstrate that the proposed method can be extended to more complex
imaging physics, we also included an anisotropic example where a subset
of the 2007 BP TTI dataset is imaged. Figure~\ref{fig:2d-bp} includes
the imaging result. From this image, we observe that all the layers are
imaged correctly and continuously. Because we have a much denser
acquisition in this case, the noise incurred by the randomized trace
estimation mostly stacks out leading to a very clean image for a
fraction of the memory cost of standard RTM. Again all calculations were
done exclusively on the GPU.

\subsection{Code availability}\label{code-availability}

Our implementation and examples are available as open-source software at
\href{:https://github.com/slimgroup/TimeProbeSeismic.jl}{TimeProbeSeismic.jl}, which extends our Julia inversion framework
\href{https://github.com/slimgroup/JUDI.jl}{JUDI.jl}\citep{witteJUDI2019}. Our code is also designed to
generalize to 3D and more complicated physics as supported by
\href{https://www.devitoproject.org}{Devito}\citep{devito-api, devito-compiler}.

\section{Discussion and conclusions}\label{discussion-and-conclusions}

We presented a proof of concept of wave-based seismic inversion with
randomized trace estimation. Aside from demonstrating that this method
leads to drastic memory reductions that allow us to form
subsurface-offset gathers while remaining on the GPU, we also showed
that the incoherent imaging artifacts mostly stack out. This allowed us
to create high fidelity images for realistic anisotropic models with
salt. Unlike existing methods based on on-the-fly Fourier transforms, we
observed that imaging artifacts induced by the random trace estimation
stack out for a high enough fold. We expect that this reliance on fold
can be relaxed when carrying sparsity-promoting imaging or full-waveform
inversion with constraints. More importantly, the drastic memory
reduction allows us to carry out the imaging exclusively on GPUs.

\section{Acknowledgement}\label{acknowledgement}

This research was carried out with the support of Georgia Research
Alliance and partners of the ML4Seismic Center.

\bibliography{bibliography}

\end{document}